\documentclass[prb, notitlepage]{revtex4-1}

\usepackage{amsmath, amssymb, amsthm, amsfonts}
\usepackage{physics}
\usepackage{graphicx}
\usepackage{float}
\usepackage{url}
\usepackage[colorlinks=false]{hyperref}
\usepackage[font={small,it}]{caption}
\usepackage{xcolor}
\usepackage{soul}

\begin{document}

\title{Constraints on the Dimensionality of Space}
\date{\today}
\author{C. A. Petway, R. D. Orlando,  A. M. McNamara, E. A. Zweig,  B. C. Caminada, E. J. Kincaid, C. V. Landgraf, C. M. Mohs, M. L. Schiff, E. Fischbach}

\email{ephraim@purdue.edu (corresponding author)}

\affiliation{Department of Physics  and Astronomy, Purdue University, West Lafayette, IN 47907, USA}

\begin{abstract}
Complex structures can only form in a universe that allows for bound states. While this is clearly observed in three-dimensions, added degrees of freedom in a higher-dimensional space preclude the immediate assumption that binding potentials can in fact exist. In this paper, we derive a constraint on the dimensionality of a universe in the presence of an arbitrary set of forces. We then apply this constraint to systems with several example potentials. In doing so, we find that bound states in higher than 3 dimensions are in fact possible under specific circumstances which we characterize. One implication of this work is that certain anomalous phenomena, such as light from stars/galaxies apparently emerging/disappearing could be evidence of extra spatial dimensions.
\end{abstract}

\maketitle

\setlength{\parindent}{0cm}
\section{Introduction}

In order to have a universe of atoms and stars, forces need to allow for bound states to occur. Without the attraction supplied by fundamental forces, the universe would be a sea of unbound particles. However, attractive forces do exist in our universe, allowing for quarks to form baryons, and eventually for atoms, stars, and galaxies to form.

The local observable universe has three spatial dimensions. On a larger scale, however, the larger universe could exhibit four or more spatial dimensions, at least in some local regions. One can explore this possibility by searching for anomalous phenomena. These could include searching for stars/ galaxies apparently emerging/ disappearing, as might happen if light from these objects passes from a region of higher dimensionality to a lower one, as well as searching for the higher dimensional phenomena which will be described in this paper. 

At the microscopic scale the dimensionality of space could also be connected to the question of the quantization of space, a question which has been considered by a number of authors over the years.[1-5] Since altering the dimensionality of space affects many equations derived in our local universe, the instability provided by increasing spatial degrees of freedom precludes the natural assumption that bound states can also occur in regions with more than three spatial dimensions. In Ref. [6], Rabinowitz demonstrated that the existence of gravitational and electromagnetic forces demands that our observable universe be only three-dimensional. Separately, L$\ddot{\text{a}}$mmerzahl and Macias have presented general arguments why the dimensionality of space-time should be four [7]. In what follows we expand upon Rabinowitz's work by deriving a constraint on a higher dimensional universe from a potential with arbitrary form. We show specifically that combinations of electromagnetic, gravitational, and short range Yukawa forces can in fact produce bound states in a Euclidean universe with greater than three spatial dimensions.
\section{Deriving the Constraint}

\subsection{Hypersphere Surface Area}

The $r$-dependence of typical potentials results from applying Gauss' law over some volume containing the source of a force field. An integral part of generalizing this derivation to $n$ dimensions is determining the surface area of an $n$-sphere. An easy way to do this is to first consider the volume $\mathcal{V}_n$ of an $n$-dimensional Euclidean sphere ($n$-sphere). A derivation of these volumes is rather involved, but has been carried out by Gipple in Ref. [8]. He finds that 
\begin{equation}\label{n_volume}
	\mathcal{V}_n=\frac{2\pi^{n/2} (r_n)^n}{n\Gamma(n/2)},
\end{equation}
where $r_n$ is the radius of the $n$-sphere, and $\Gamma$ is the usual gamma function. To obtain the surface area $A_{n-1}$ of the $n$-sphere, we differentiate the volume with respect to $r_n$, which yields,
\begin{equation}\label{n_area}
	A_{n-1} = \frac{d \mathcal{V}_n}{dr_n}=\frac{2\pi^{n/2}(r_n)^{n-1}}{\Gamma(n/2)}.
\end{equation}

\subsection{General Potential}

Rabinowitz provided a constraint on the number of spatial dimensions a universe can have while maintaining the ability to form bound states [6]. In his paper, he used the Virial Theorem for a Coulomb potential and found that bound orbits are only possible in 3-dimensional universes. If, however, one assumes that potentials can take on many different functional forms, other solutions appear. Consider to start with a 3-dimensional Coulomb or gravitational potential given by
\begin{equation}\label{3_potential_coul}
	V_3 = -\frac{K/q}{4\pi r_3},
\end{equation}
where $K = Ze^2/\epsilon_0$ for the Coulomb force, or $4\pi GMm$ for the gravitational force, and $q$ is the generalized charge of the object the potential is acting on. Specifically, $q=e$ for the electromagnetic force, or $q=m$ for the gravitational force. The force is then given by the negative gradient of such a potential when multiplied by the generalized charge $q$,
\begin{equation}\label{3_force_coul}
	\vec{F}_3 = (-\vec{\nabla}V_3)q = -\frac{K}{4\pi (r_3)^2}\hat{r}_3.
\end{equation} 
To generalize this result to an arbitrary functional form in higher Euclidean spatial dimensions, we start by multiplying (\ref{3_potential_coul}) by an appropriate function of interest, $\phi(r_n)$, and take the gradient of this new potential. Noting that $A_2=4\pi (r_3)^2$ from (\ref{n_area}), we find
\begin{equation}\label{3_potential_gen}
	V_3 = -\frac{K/q}{4\pi r_3}\phi(r_3),
\end{equation}
\begin{equation}\label{3_force_gen}
	\vec{F}_3 = (-\vec{\nabla}V_3)q = -\frac{K}{A_2}\left[\phi(r_3)-r_3\phi'(r_3)\right]\hat{r_3},
\end{equation}
where $\phi'$ denotes $d\phi(r_3)/dr_3$. To check that this maintains the correct relationship between the force and the potential, we can take $\phi(r_3)=1$ in (\ref{3_force_gen}), which evidently reproduces (\ref{3_force_coul}) as expected. By multiplying both sides of (\ref{3_force_gen}) by $A_2$, we recover Gauss' law,
\begin{equation}\label{3_gauss}
	\oint_{A_{2}} \vec{F}_3 \cdot d\vec{A}_2 = -K\left[\phi(r_3)-r_3\phi'(r_3)\right].
\end{equation}

\subsection{$n$-Dimensional Forces}
Since the expression for the surface area of a sphere in any dimension is known, we can extend the relation in (\ref{3_gauss}) to an arbitrary dimension,
\begin{equation}\label{n_gauss}
	\oint_{A_{n-1}} \vec{F}_n\cdot d\vec{A}_{n-1} = -K\left[\phi(r_n)-r_n\phi'(r_n)\right].
\end{equation}
The force $\vec{F}_n$ will always be normal to the Gaussian surface in a Euclidean space, and hence the right hand side of (\ref{n_gauss}) simplifies to the product of the force and the surface area. Solving for $\vec{F}_n$ leads to the generalized force in an arbitrary dimension,
\begin{equation}\label{n_force_gen}
	\vec{F}_n = -\frac{K}{A_{n-1}}\left[\phi(r_n)-r_n\phi'(r_n)\right]
	=-\frac{K\Gamma(n/2)}{2\pi^{n/2}r_n^{n-1}}[\phi(r_n)-r_n\phi'(r_n)]\hat{r}_n.
\end{equation}

\subsection{Existence of Bound States in Higher Dimensions}
To simplify the problem, we focus on the case of circular orbits, since for any elliptical orbit there will be a circular orbit of the same energy. For an object to be bound while in orbit, its total energy must be negative,
\begin{equation}\label{tot_E}
	E_n = T_n + U_n < 0,
\end{equation}
where $T_n$ is the kinetic energy and $U_n$ is the potential energy. For a force $\vec{F}_n$ to maintain an object in a circular orbit, the magnitude of the force $\abs{\vec{F}_n}$ must be given by $m(v_n)^2/r_n$. Combining this with the kinetic energy $T_n = \frac{1}{2}m(v_n)^2$, it follows that
\begin{equation}\label{n_kinetic_gen}
	T_n = \frac{1}{2}\abs{F_n}r_n=\abs{\frac{K\Gamma(n/2)}{4\pi^{n/2}(r_n)^{n-2}}\left[\phi(r_n)-r_n\phi'(r_n)\right]}.
\end{equation}
The potential $U_n$ is given by the path integral of the force,
\begin{equation}\label{n_potential_gen}
	U_n = -\int_{\infty}^{r_n}\vec{F}_n\cdot d(\vec{r}_n)'
	=\frac{K\Gamma(n/2)}{2\pi^{n/2}}\left[\int_\infty^{r_n}\frac{\phi(x)}{x^{n-1}}dx-\int_\infty^{r_n}\frac{\phi'(x)}{x^{n-2}}dx\right].
\end{equation}
Using (\ref{n_kinetic_gen}) and (\ref{n_potential_gen}), a constraint on $n$ and $\phi(r_n)$ can be found by looking for solutions which satisfy (\ref{tot_E}),
\begin{equation}\label{n_constraint_gen_abs}
	\frac{1}{2}\abs{\frac{\phi(r_n)}{(r_n)^{n-2}}-\frac{\phi'(r_n)}{(r_n)^{n-3}}}
	+\int_\infty^{r_n}\frac{\phi(x)}{x^{n-1}}dx-\int_\infty^{r_n}\frac{\phi'(x)}{x^{n-2}}dx<0.
\end{equation}
For the force to be net attractive, the expression within the absolute value must be positive, for if it were negative, the force would be repulsive and bound states would not be possible. This leads to the following set of constraints on the forces,
\begin{equation}\label{phi_constraint}
	\phi(r_n)\geq r_n\phi'(r_n),
\end{equation}
\begin{equation}\label{n_constraint_gen}
	\frac{1}{2}\left(\frac{\phi(r_n)}{(r_n)^{n-2}}-\frac{\phi'(r_n)}{(r_n)^{n-2}}\right)+\int_\infty^{r_n}\frac{\phi(x)}{x^{n-1}}dx-\int_\infty^{r_n}\frac{\phi'(x)}{x^{n-2}}dx<0.
\end{equation}

\section{Applying the Constraint}
Here we consider several different expressions for $\phi(r_n)$ for which we derive the specific constraint.

\subsection{Coulomb Potential}
 We begin with a Coulomb potential which characterizes both the gravitational and electromagnetic forces. Although these forces are typically referred to as inverse square law forces, this name loses its descriptive meaning when generalized to $n$ dimensions. This potential is a convenient starting point because this case has already been analyzed by Rabinowitz in Ref. [6]. In the notation of (\ref{3_potential_gen}) and (\ref{3_force_gen}), $\phi(r_n)$ and its derivative are then given by
\begin{equation}\label{phi_coul}
	\phi(r_n) = 1
\end{equation}
\begin{equation}\label{phi_p_coul}
	\phi'(r_n) = 0.
\end{equation}
Clearly these functions satisfy (\ref{phi_constraint}). Combining these two values and (\ref{n_constraint_gen}) gives the following inequality,
\begin{equation}\label{coul_cons}
	0>\frac{1}{2}\left(\frac{1}{r^{n-2}_n}\right)+\int_\infty^{r_n}\frac{1}{x^{n-1}}dx.
\end{equation}
It follows that the only physically meaningful value of $n$ consistent with (\ref{coul_cons}) is $n=3$ as shown by Rabinowitz.

\subsection{Yukawa Potential}
\begin{figure}[H]
	\centering
    \includegraphics[scale=1]{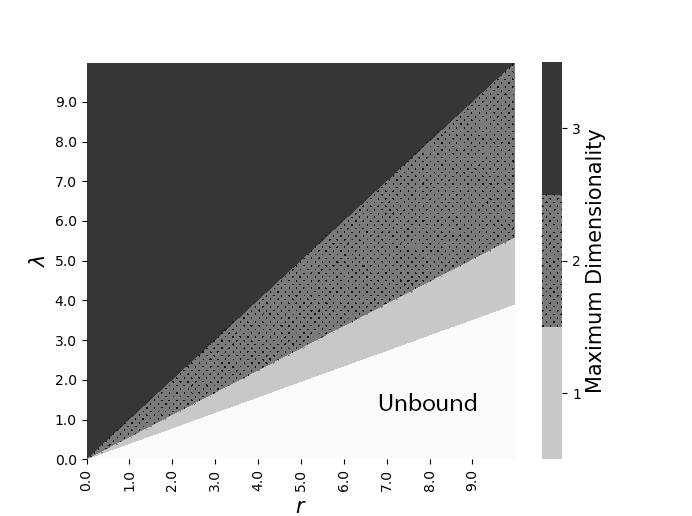}
    \caption{Plot of the maximum dimension (shaded region) allowing bound states to exist for a Yukawa potential. Each bound region allows for bound states from 1 dimension up to its maximum dimensionality. Bound states can only arise in 3 or fewer dimensions, and then only for separations shorter than the range $\lambda$ of the force in three dimensions, and some multiple of $\lambda$ for one and two dimensions. The region where bound states cannot arise for any dimensionality is labeled "Unbound."}
    \label{yuk_fig}
\end{figure}

The strong nuclear force can be described by a Yukawa potential, i.e., $V=-g^2e^{-r/\lambda}/r$ where $\lambda$ is a constant. This can be analyzed in our framework by assuming the following attractive expression for $\phi(r_n)$,
\begin{equation}\label{phi_yuk}
	\phi(r_n) = e^{-r_n/\lambda},
\end{equation}
\begin{equation}\label{phi_p_yuk}
	\phi'(r_n) = - \frac{e^{-r_n/\lambda}}{\lambda}.
\end{equation}
Applying (\ref{phi_constraint}) to these functions gives $1\geq-r_n/\lambda$. Since $r_n$ and $\lambda$ are positive by definition, this condition is always satisfied. Thus the constraint from (\ref{n_constraint_gen}) becomes
\begin{equation}
	0>\frac{1}{2}\left(\frac{e^{-r_n/\lambda}}{r_n^{n-2}}+\frac{e^{-r_n/\lambda}}{\lambda r^{n-3}_n}\right)+\int_\infty^{r_n}\frac{e^{-x/\lambda}}{x^{n-1}}dx+\int_\infty^{r_n}\frac{e^{-x/\lambda}}{\lambda x^{n-2}}dx.
\end{equation}
The contribution from $e^{-r_n/\lambda}/r_n^{n-2}$ can be factored out of the kinetic energy term, and $-1/\lambda^{n-2}$ can be factored out of the potential energy term, while inverting the bounds of the integral. Doing so leads to
\begin{equation}
	0>\frac{e^{-r_n/\lambda}}{2r_n^{n-2}}\left(1+\frac{r_n}{\lambda}\right)
	-\frac{1}{\lambda^{n-2}}\left[\int_{r_n}^\infty\frac{\lambda^{n-2}e^{-x/\lambda}}{x^{n-1}}dx+\int_{r_n}^\infty \frac{\lambda^{n-3}e^{-x/\lambda}}{x^{n-2}}dx\right].
\end{equation}
We next substitute $u=x/\lambda$ and $du=dx/\lambda$, while multiplying all terms by $2r_n^{n-2}$. This leads to the inequality,
\begin{equation}\label{pre_gamma}
	0>e^{-r_n/\lambda}\left(1+\frac{r_n}{\lambda}\right)-2\left(\frac{r_n}{\lambda}\right)^{n-2}\left[\int_{r_n/\lambda}^\infty u^{1-n}e^{-u}du+\int_{r_n/\lambda}^\infty u^{2-n}e^{-u}du\right].
\end{equation}
Each of the integrals can be expressed in terms of an upper incomplete gamma function, $\Gamma(a,b)$, defined as
\begin{equation}\label{gamma_func}
	\Gamma(a,b)=\int_{b}^{\infty}x^{a-1}e^{-x}dx.
\end{equation}
Combining (\ref{gamma_func}) and (\ref{pre_gamma}) gives
\begin{equation}\label{post_gamma}
	0>e^{-r_n/\lambda}\left(1+\frac{r_n}{\lambda}\right)-2\left(\frac{r_n}{\lambda}\right)^{n-2}\left[\Gamma\left(2-n,\frac{r_n}{\lambda}\right)+\Gamma\left(3-n, \frac{r_n}{\lambda}\right)\right].
\end{equation}
Upper incomplete gamma functions have the following property, which is useful at this point,
\begin{equation}\label{gamma_prop}
	\Gamma(a+1, b)=a\Gamma(a,b)+ b^{a}e^{-b}.
\end{equation}
Combining (\ref{post_gamma}) and (\ref{gamma_prop}) leads to
\begin{equation}
	0>e^{-r_n/\lambda}\left(1+\frac{r_n}{\lambda}\right)-2\left(\frac{r_n}{\lambda}\right)^{n-2}\left[\left(\frac{r_n}{\lambda}\right)^{2-n}e^{-r_n/\lambda}+(3-n)\Gamma\left(2-n,\frac{r_n}{\lambda}\right)\right].
\end{equation}
Simplifying this further yields the following constraint,
\begin{equation}
	0>e^{-r_n/\lambda}\left(\frac{r_n}{\lambda}-1\right)-(6-2n)\left(\frac{r_n}{\lambda}\right)^{n-2}\Gamma\left(2-n,\frac{r_n}{\lambda}\right).
\end{equation}
This constraint is graphically represented in Fig. \ref{yuk_fig}. In this figure it is clear that a Yukawa force can only bind particles in three or fewer dimensions. Additionally, there is a maximum range over which these states can occur, which is $\lambda$ in three dimensions and a multiple of $\lambda$ in one and two dimensions.

\subsection{Coulomb$(1+\text{constant}/r)$ Potential}
\begin{figure}[H]
	\centering
    \includegraphics[scale=1]{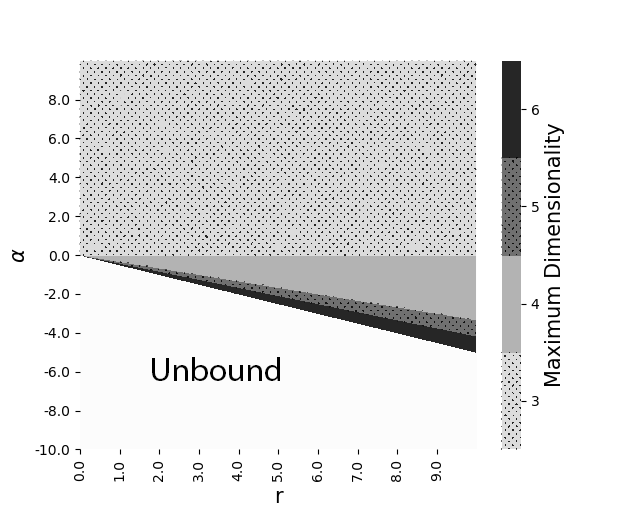}
    \caption{Plot of the maximum dimension allowing for bound states of a system with both a Coulomb potential and a potential that decreases faster by a factor of $r^{-1}$. Each bound region allows for bound states from 3 dimensions up to its maximum dimensionality. When the forces oppose each other, there exists a small band of distances in which there are bound states in higher spatial dimensions. For computational reasons, six spatial dimensions were the highest tested. However, bound states do exist in higher dimensions. The region where bound states cannot arise for any dimensionality is labeled "Unbound."}
    \label{coul_inv_fig}
\end{figure}

We can also consider potentials $\phi(r_n)$ that have more than one term. In this case, one of the forces decreases with distance faster than the other by a factor of $1/(r_n)$. There are  simple expressions for $\phi(r_n)$ and $\phi'(r_n)$ which we can use to explore this possibility:
\begin{equation}\label{phi_coul_inv}
	\phi(r_n) = 1 + \frac{\alpha}{r_n},
\end{equation}
\begin{equation}\label{phi_p_coul_inv}
	\phi'(r_n) = - \frac{\alpha}{r_n^2}.
\end{equation}
Here $\alpha$ denotes the ratio of the strength of the faster decaying force to the slower one at $r_n=1$, and determines whether it contributes an attractive (positive) or repulsive (negative) force to the attractive Coulomb force. Using (\ref{phi_constraint}), we can see that there is a lower bound on the value of $\alpha$,
\begin{equation}\label{alpha_cond}
	1+\frac{\alpha}{r_n}\geq r_n(-\frac{\alpha}{r_n^2}).
\end{equation}
It follows from (\ref{alpha_cond}) that the lower bound on $\alpha$ is given by
\begin{equation}\label{coul_inv_a_const}
	\alpha\geq-r_n/2.
\end{equation}
Inserting this value of $\alpha$ into $\phi(r_n)$ and $\phi'(r_n)$, and combining with (\ref{n_constraint_gen}) leads to the following constraint on $n$,
\begin{equation}\label{coul_inv_init_cons}
	0>\frac{1}{2}\left(\frac{1+\alpha/r_n}{r_n^{n-2}}+\frac{\alpha/r_n^2}{r_n^{n-3}}\right)+ \int_\infty^{r_n}\frac{1+\alpha/x}{x^{n-1}}dx + \int_\infty^{r_n}\frac{\alpha/x^2}{x^{n-2}}dx.
\end{equation}
By separating the first integral into two contributions, we note that the second has the same functional form as the last integral. Combining the various terms in (\ref{coul_inv_init_cons}) we find,
\begin{equation}
	0>\frac{1}{2}\left(\frac{1}{r_n^{n-2}} + \frac{\alpha}{r_n^{n-1}}+\frac{\alpha}{r_n^{n-1}}\right)+\int_\infty^{r_n}\frac{1}{x^{n-1}}dx+2\int_\infty^{r_n}\frac{\alpha}{x^n}dx.
\end{equation}
Assuming again that $n>1$ we can perform these integrations which gives,
\begin{equation}
	0>\frac{1}{2}\left(\frac{1}{r_n^{n-2}}+\frac{2\alpha}{r_n^{n-1}}\right)-\frac{1}{(n-2)r_n^{n-2}}-\frac{2\alpha}{(n-1)r_n^{n-1}}.
\end{equation}
After multiplying through by $2/r_n^{n-2}$, we find
\begin{equation}
	0>1+\frac{2\alpha}{r_n}-\frac{2}{n-2}-\frac{4\alpha}{(n-1)r_n},
\end{equation}
and hence,
\begin{equation}\label{coul_inv_n_const}
	0>\frac{n-4}{n-2}+\frac{2\alpha}{r_n}\left(\frac{n-3}{n-1}\right).
\end{equation}
The application of both equations (\ref{coul_inv_a_const}) and (\ref{coul_inv_n_const}) is graphically represented in Figure \ref{coul_inv_fig}. 

\setlength{\parindent}{0cm}
It can be seen in this figure that the maximum dimensionality allowing for bound states increases as the parameters approach the lower bound of $\alpha$. If we take $\alpha$ to be at its minimum possible value for a given $r_n$, that is $\alpha=-r_n/2$, and apply this to (\ref{coul_inv_n_const}), we find
\begin{equation}\label{coul_inv_inf_cond}
	\frac{n-3}{n-1}>\frac{n-4}{n-2}.
\end{equation}
Since (\ref{coul_inv_inf_cond}) is equivalent to the statement that $6>4$, it follows that at this minimum value of $\alpha$, any number of spatial dimensions are allowed.

\subsection{Coulomb plus Yukawa Potential}
We next consider a more physical form of a potential with two terms, namely the sum of a Yukawa and a Coulomb potential. In fact, we have actual examples of particles that interact via forces of both kinds, such as quarks. In the case of an attractive Coulomb force, the $\phi(r_n)$ and $\phi'(r_n)$ are then as follows,
\begin{equation}\label{phi_coul_yuk}
	\phi(r_n) = 1+\Omega e^{-r_n/\lambda},
\end{equation}
\begin{equation}\label{phi_p_coul_yuk}
	\phi'(r_n) = -\frac{\Omega}{\lambda}e^{-r_n/\lambda}.
\end{equation}
Here $\Omega$ is the ratio of the strengths of the Yukawa and Coulomb potentials and its sign determines whether the Yukawa potential is attractive (positive) or repulsive (negative), while $\lambda$ is the range of the Yukawa force. Again we have a constraint on $\Omega$ from (\ref{phi_constraint}),
\begin{equation}
	1+\Omega e^{-r_n/\lambda}\geq r_n\left(-\frac{\Omega}{\lambda}e^{-r_n/\lambda}\right),
\end{equation}
which is equivalent to
\begin{equation}\label{omega_cons}
	\Omega e^{-r_n/\lambda}(1+\frac{r_n}{\lambda})\geq -1.
\end{equation}
It follows that (\ref{omega_cons}) implies a lower bound the value of $\Omega$ given by
\begin{equation}\label{coul_yuk_O_const}
	\Omega \geq -\frac{e^{r_n/\lambda}}{1+r_n/\lambda}.
\end{equation}
A constraint on $n$ from (\ref{n_constraint_gen}) can be obtained by using the expression for $\phi(r_n)$ and $\phi'(r_n)$ for this potential:
\begin{equation}\label{init_ineq}
	0>\frac{1}{2}\left(\frac{1+\Omega e^{-r_n/\lambda}}{(r_n)^{n-2}}+\frac{\Omega e^{-r_n/\lambda}}{\lambda (r_n)^{n-3}}\right)+\int_\infty^{r_n}\frac{1+\Omega e^{-x/\lambda}}{x^{n-1}}dx+\int_\infty^{r_n}\frac{\Omega e^{-x/\lambda}}{\lambda x^{n-2}}dx.
\end{equation}
The first of the integrals can be separated into two contributions, the first of which is simple to evaluate. After doing so, $\Omega/\lambda^{n-2}$ can be factored out of the remaining integrals and $r_n^{2-n}/2$ can be factored out of the kinetic energy term. The inequality in (\ref{init_ineq}) then assumes the form
\begin{equation}
\begin{split}
	0>&\frac{r_n^{2-n}}{2}\bigg[1+\Omega e^{-r_n/\lambda}\left(1+\frac{r_n}{\lambda}\right)\bigg]-\frac{r_n^{2-n}}{n-2}+\frac{\Omega}{\lambda^{n-2}}\bigg[\int_\infty^{r_n}\frac{\lambda^{n-2}e^{-x/\lambda}}{x^{n-1}}dx\\
	&+ \int_\infty^{r_n}\frac{\lambda^{n-3}e^{-x/\lambda}}{x^{n-2}}dx\bigg].
\end{split}
\end{equation}
Substituting $u=x/\lambda$ and $du=dx/\lambda$, we find
\begin{equation}
\begin{split}
	0>& 1-\frac{2}{n-2}+\Omega e^{-r_n/\lambda}\left(1+\frac{r_n}{\lambda}\right)
	+2\Omega\left(\frac{r_n}{\lambda}\right)^{n-2}\bigg[\int_\infty^{r_n/\lambda}u^{1-n}e^{-u}du\\
	&+\int_\infty^{r_n/\lambda}u^{2-n}e^{-u}du\bigg].
\end{split}
\end{equation}
The first two terms of the previous equation simplify to the expression for the constraint on the Coulomb potential given in (\ref{coul_cons}). Again these integrals are of the same form as the upper incomplete gamma function given in (\ref{gamma_func}), and hence,
\begin{equation}
\begin{split}
	0>&\frac{n-4}{n-2}+\Omega e^{-r_n/\lambda}\left(1+\frac{r_n}{\lambda}\right)
	-2\Omega\left(\frac{r_n}{\lambda}\right)^{n-2}\bigg[\Gamma\left(2-n, \frac{r_n}{\lambda}\right)\\
	&+\Gamma\left(3-n, \frac{r_n}{\lambda}\right)\bigg].
\end{split}
\end{equation}

\begin{figure}[H]
	\centering
    \includegraphics[scale=1]{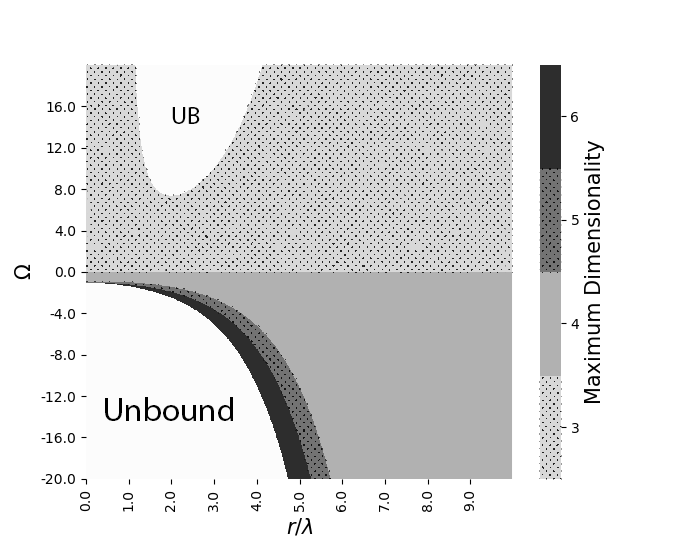}
    \caption{Plot of the maximum dimension allowed for bound states of a system in the presence of the sum of a Coulomb potential and a Yukawa potential. Each bound region allows for bound states from 3 dimensions up to its maximum dimensionality. We find two unbounded regions, where the upper one (labeled "UB") is due to the finite range of the Yukawa potential, and the lower (labeled "Unbound") is due to the net force being repulsive. As the parameters approach this lower unbounded region, the maximum dimension allowing for bound states significantly increases. For computational reasons, six spatial dimensions were the highest tested. However, bound states do exist in higher dimensions.}
    \label{coul_yuk_fig}
\end{figure}

We can also use the property of incomplete gamma functions given in (\ref{gamma_prop}) to derive the following inequality:
\begin{equation}\label{half_simp}
\begin{split}
	0>&\frac{n-4}{n-2}+\Omega e^{-r_n/\lambda}\left(1+\frac{r_n}{\lambda}\right)
	-2\Omega\left(\frac{r_n}{\lambda}\right)^{n-2}\bigg[\left(\frac{r_n}{\lambda}\right)^{2-n}e^{-r_n/\lambda}\\
	&+(3-n)\Gamma\left(2-n,\frac{r_n}{\lambda}\right)\bigg].
\end{split}
\end{equation}
The inequality in (\ref{half_simp}) simplifies to the following constraint,
\begin{equation}
	0>\frac{n-4}{n-2}+\Omega e^{-r_n/\lambda}\left(\frac{r_n}{\lambda}-1\right)-\Omega(6-2n)\left(\frac{r_n}{\lambda}\right)^{n-2}\Gamma\left(2-n,\frac{r_n}{\lambda}\right).
\end{equation}
Unlike the case for the Coulomb potential, this constraint allows for bound states in higher than 3 dimensions but only for specific values of $\Omega$ and $\lambda$. This is graphically represented in Figure \ref{coul_yuk_fig}.

\setlength{\parindent}{0cm}
In the $\Omega>0$ section of this plot, we see an unbound region that is between two bounded regions. Physically, this represents a configuration of forces where there is binding up until some distance $r\approx\lambda$ and then binding again at some longer distance $r=c\lambda$ where $c$ is some constant depending on $\Omega$. This is because (as with the case of only the Yukawa potential) the Yukawa term is no longer capable of binding a particle with a displacement greater than its range $\lambda$. However the Yukawa potential is still strong enough to accelerate the particle such that its kinetic energy cannot be bound by the Coulomb potential. After a certain distance (depending on the ratio of the strengths of the forces) the strength of the Yukawa potential is decreased by its exponential term so much that it is negligible compared to the infinite range Coulomb potential. Thus binding is allowed again. We see in the $\Omega<0$ section of this plot similar behavior to the case of the Coulomb($1+\text{constant}/r)$ potential in that the maximum dimensionality increases as the parameters approach the lower bound. It thus appears that we might also see the possibility of near infinite dimensions at this minimum $\Omega$. The addition of a repulsive Yukawa potential allows for these higher dimensional bound states because it weakens the acceleration, and thus the kinetic energy, far more than it weakens the potential energy. In higher dimensions, the increased degrees of freedom cause the kinetic energy of the particle to increase and the potential to decrease faster with respect to distance, however the presence of a repulsive potential of a different functional form suppresses this effect to the extent that binding is made possible again. Noting that since the kinetic energy term vanishes at the minimum $\Omega$, we can derive the following constraint from (\ref{n_constraint_gen}), using $\phi(r_n)$, and $\phi'(r_n)$ for this potential:
\begin{equation}
	0>\int_\infty^{r_n}\frac{1+\Omega e^{-x/\lambda}}{x^{n-1}}dx+\int_\infty^{r_n}\frac{\Omega e^{-x/\lambda}}{\lambda x^{n-2}}dx.
\end{equation}
As before, we can separate the first integral into two, the first of which can be easily evaluated. After removing a factor an $\Omega/\lambda^{n-2}$ from the remaining integrals, we find
\begin{equation}
	0>-\frac{r_n^{2-n}}{n-2}+\frac{\Omega}{\lambda^{n-2}}\left[\int_\infty^{r_n}\frac{\lambda^{n-2}e^{-x/\lambda}}{x^{n-1}}dx + \int_\infty^{r_n}\frac{\lambda^{n-3}e^{-x/\lambda}}{x^{n-2}}dx\right].
\end{equation}
As we have done previously, we can express the two remaining integrals as an upper incomplete gamma function, leading to,
\begin{equation}\label{unsimplified_constraint}
	0>-\frac{r_n^{2-n}}{n-2}-\frac{\Omega}{\lambda^{n-2}}\left[\Gamma\left(2-n,\frac{r_n}{\lambda}\right)+\Gamma\left(3-n,\frac{r_n}{\lambda}\right)\right].
\end{equation}
The expression in (\ref{unsimplified_constraint}) can be simplified using (\ref{gamma_prop})  yielding,
\begin{equation}
	0>-\frac{1}{n-2}-\Omega\left(\frac{r_n}{\lambda}\right)^{n-2}\left[e^{-r_n/\lambda}\left(\frac{r_n}{\lambda}\right)^{2-n}+(3-n)\Gamma\left(2-n,\frac{r_n}{\lambda}\right)\right].
\end{equation}
Simplifying further gives an expression similar to the final constraint found before,
\begin{equation}
	0>-\frac{1}{n-2}-\Omega e^{-r_n/\lambda}-(3-n)\Omega\left(\frac{r_n}{\lambda}\right)^{n-2}\Gamma\left(2-n, \frac{r_n}{\lambda}\right).
\end{equation}
Substituting in the minimum value of $\Omega$ given in (\ref{coul_yuk_O_const}) we find,
\begin{equation}
	0>-\frac{1}{n-2}+\frac{1}{1+r_n/\lambda}+\frac{3-n}{1+r_n/\lambda}\left(\frac{r_n}{\lambda}\right)^{n-2}e^{r_n/\lambda}\Gamma\left(2-n,\frac{r_n}{\lambda}\right).
\end{equation}
Next we can multiply through by $1/(1+r_n/\lambda)$ which gives,
\begin{equation}
	0>-\frac{1+r_n/\lambda}{n-2}+1+(3-n)\left(\frac{r_n}{\lambda}\right)^{n-2}e^{r_n/\lambda}\Gamma\left(2-n,\frac{r_n}{\lambda}\right)\\.
\end{equation}
Let us now consider the limit of this expression as $n \rightarrow \infty$. The first term vanishes for all finite $r_n$. For simplicity we substitute $x=r_n/\lambda$, which gives the following constraint for bound states to exist in any number of dimensions,
\begin{equation}\label{n_lim_inf}
	0> \lim_{n\rightarrow\infty}\left[1+(3-n)x^{n-2}e^{x}\Gamma(2-n,x)\right].
\end{equation}
The expression in (\ref{n_lim_inf}) can be simplified by expressing the inequality in the form
\begin{equation}\label{lim_gamma}
	1<\lim_{n\rightarrow\infty}\left[nx^{n-2}e^{x}\Gamma(2-n,x)\right].
\end{equation}
If the gamma function is then re-expressed in an integral form this becomes,
\begin{equation}\label{lim_int}
	1<\lim_{n\rightarrow\infty}\left[nx^{n-2}e^{x}\int_x^{\infty}u^{1-n}e^{-u}du\right].
\end{equation}
Although we will now show that this constraint is not satisfied as $n\rightarrow\infty$, the limit of this expression does yield an interesting result. Integrating by parts the integral in (\ref{lim_int}) gives,
\begin{equation}
\begin{split}
    \int_{x}^{\infty}y^{1-n}e^{-y}dy
    &=\eval{\left(\frac{y^{2-n}}{2-n}e^{-y}\right)}^\infty_x+\frac{1}{2-n}\int_{x}^{\infty}{y^{2-n}e^{-y}dy}\\
    &=-\frac{x^{2-n}}{2-n}e^{-x}+\ \frac{1}{2-n}\int_{x}^{\infty}{y^{2-n}e^{-y}dy}.
\end{split}
\end{equation}
Continuing to integrate by parts yields,
\begin{equation}\label{expanded_int}
    \int_{x}^{\infty}{y^{1-n}e^{-y}dy}=-\frac{x^{2-n}}{2-n}e^{-x}\left(1+\frac{x}{3-n}+\frac{x^2}{\left(3-n\right)\left(4-n\right)}+\ldots\right).
\end{equation}
Taking the limit as $n\rightarrow\infty$ in (\ref{lim_gamma}), and retaining only the leading term in the expanded integral shown in (\ref{expanded_int}), we find  

\begin{equation}
\begin{split}
    \lim_{n\rightarrow\infty}{\frac{\Gamma\left(2-n,x\right)}{\frac{1}{n}x^{2-n}e^{-x}}}
    &=\left. \lim_{n\rightarrow\infty}{-\frac{x^{2-n}}{2-n}e^{-x}\left(1+\frac{x}{3-n}+\frac{x^2}{\left(3-n\right)\left(4-n\right)}+\ldots\right)} \middle/ {\left(\frac{1}{n}x^{2-n}e^{-x}\right)} \right.\\
    &=\left. \lim_{n\rightarrow\infty}{\left(\frac{x^{2-n}}{n-2}e^{-x}\right)} \middle/ {\left( \frac{1}{n}x^{\left(2-n\right)}e^{-x}\right) }\right.=1.
\end{split}
\end{equation}

It follows that the right hand side of (\ref{lim_gamma}) approaches 1 as $n\rightarrow\infty$, at which point  (\ref{lim_gamma}) is no longer satisfied. However, we observe bound states in increasingly high dimensions just before this constraint is violated, which implies that there should be bound states in all finite dimensions as the parameters approach this lower bound. As a specific example, we find that the constraint for $\Omega=-2002.405981$ and $r_n/\lambda=10$ is $n\leq324350$.

\begin{figure}[H]
	\centering
	\includegraphics[scale=1]{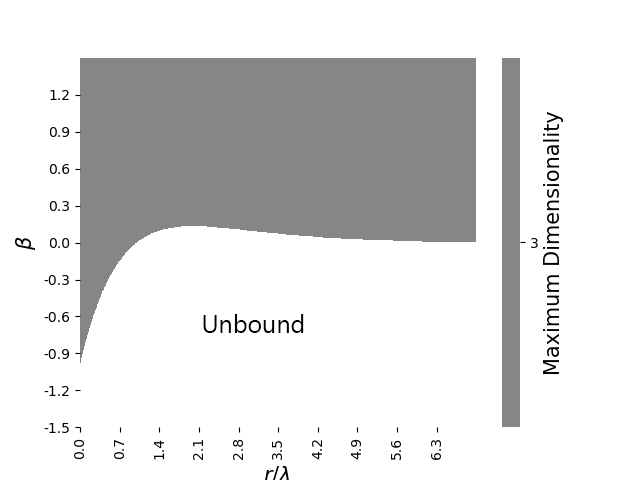}
	\caption{Plot of the maximum dimension allowed for bound states of a system in the presence of the sum of a Coulomb potential and an attractive Yukawa potential. The bound region allows for bound states only in 3 dimensions. The bound region in the $\beta<0$ section of the graph is due to the repulsive Coulomb potential being too weak to prevent the binding of the Yukawa potential, and the unbound region of the $\beta>0$ section of the graph corresponds to the "UB" region from Figure \ref{coul_yuk_fig}.}
	\label{yuk_coul_fig}
\end{figure}

Thus far the Coulomb potential has been constrained to be attractive while the Yukawa potential can be either repulsive or attractive. The constraint derived in this paper can be easily applied to the inverse situation. If the Yukawa potential is maintained to be attractive and the Coulomb potential is allowed to be either attractive or repulsive, the functional form $\phi(r_n)$ is given by
\begin{equation}
	\phi(r_n)=e^{-r_n/\lambda}+\beta,
\end{equation}
\begin{equation}
	\phi'(r_n)=\frac{-e^{-r_n/\lambda}}{\lambda},
\end{equation}
where $\beta$ is the ratio of the strengths between the Yukawa and Coulomb potentials, and is analogous to the $\Omega$ used previously. Using a derivation very similar to the one performed earlier in this section, we find that the constraint for this configuration of forces is
\begin{equation}
	0> e^{-r_n/\lambda}\left(\frac{r_n}{\lambda}-1\right)-(6-2n)\left(\frac{r_n}{\lambda}\right)^{n-2}\Gamma\left(2-n,\frac{r_n}{\lambda}\right)+\beta\frac{n-4}{n-2}.
\end{equation}
This constraint is graphically represented in Figure \ref{yuk_coul_fig}. It can be seen that when the Yukawa potential is constrained to be attractive there can only be bound states in three dimensions. This is because the balancing of parameters required to satisfy (\ref{n_constraint_gen}) for higher values of $n$ fail to satisfy (\ref{phi_constraint}).
	
\section{Conclusions}
For the building blocks of our universe to do more than simply scatter off one another, there must be situations in which it is energetically advantageous for these to couple, and to thus form more complex structures. The constraints derived in this paper provide a simplified tool for analyzing the Euclidean spatial dimensions where such behavior can arise. The expected results for the Coulomb and gravitational interactions, as well as the Yukawa potential, can be derived from these constraints. When applied to situations where multiple forces act on the same objects, these constraints allow for more spatial dimensions than the 3 spatial dimensions one would normally expect, and the higher dimensional phenomena described in this paper provide a means of searching for regions of the universe with differing dimensionality, as well as a description of some of the physics for such regions of space. The utility of these constraints in determining the allowed dimensionality of our universe could have far ranging implications from elementary particle physics to cosmology. [14]

Although we have limited our discussion to regions of our universe which can be characterized, at least approximately, as Euclidean [9-13], it appears possible to extend these arguments to General Relativity as will be discussed elsewhere. 

\section*{Acknowledgements}
The authors wish to thank several Megan McDuffie and Ben Hayward for their contributions in the early stages of this project.

\section*{References}
[1] H. S. Snyder, Quantized Space-Time, Phys. Rev. \textbf{71} 38 (1947).

\bigskip
\noindent
[2] C. N. Yang, On Quantized Space-Time, Phys. Rev. \textbf{72} 874 (1947).

\bigskip
\noindent
[3] A. Schild, Discrete Space-Time and Integral Lorentz Transformations, Phys. Rev. \textbf{73} 414 (1978).

\bigskip
\noindent
[4] E. Fischbach, Coupling of Internal and Quantized Space-Time Symmetries, Phys. Rev. \textbf{137} B642 (1965).

\bigskip
\noindent
[5] B. Kur\c{s}uno\v{g}lu, New Symmetry Group for Elementary Particles. I.
Generalization of Lorentz Group Via Electrodynamics, Phys. Rev. \textbf{135}, B761 (1964).

\bigskip
\noindent
[6] M. Rabinowitz. Why Observable Space is Solely Three Dimensional. {Advanced Studies in Theoretical Physics}, \textbf{8} (2014), 689-700.

\bigskip
\noindent
[7] C. L$\ddot{\text{a}}$mmerzahl and A. Macias. On the Dimensionality of Space-time. {Journal of Mathematical Physics}, \textbf{34} (1993), 4540-4553.

\bigskip
\noindent
[8] J. Gipple. The Volume of n-Balls. {Rose-Hulman Undergraduate Mathematics Journal}, \textbf{15} (2014), 238-248.

\bigskip
\noindent
[9] W. D. McGlinn, Problem of Combining Interaction Symmetries and Relativistic Invariance,  Phys. Rev. Letters \textbf{12}, 467 (1964).

\bigskip
\noindent
[10] F. Coester, M. Hamermesh, and W. D. McGlinn, Internal Symmetry and Lorentz Invariance, Phys. Rev. \textbf{135}, B451 (1964) .

\bigskip
\noindent
[11] A. O. Barut, Dynamical Symmetry Group Based on Dirac Equation and Its Generalization to Elementary Particles,  Phys. Rev. \textbf{135}, B839 (1964).

\bigskip
\noindent
[12] O. W. Greenberg, Coupling of Internal and Space-Time Symmetries, Phys. Rev. \textbf{135}, B 1447 (1964).

\bigskip
\noindent
[13] M. E. Mayer, H. J. Schnitzer, E. C. G. Sudarshan, R. Acharya, and M. Y. Han, "Concerning Space-Time and Symmetry Groups," Phys. Rev. \textbf{136}, B888 (1964).

\bigskip
\noindent
[14] Bo Qin, Ue-Li Pen, Joseph Silk, "Observational Evidence for Extra Dimensions from Dark Matter", arxiv:astro-ph/0508572v1 (2005).
\end{document}